\documentclass{article}
\usepackage{fullpage}
\usepackage[utf8]{inputenc}
\usepackage[all]{xy}
\usepackage{amssymb}
\usepackage{amsmath}
\usepackage{enumerate}
\usepackage{graphicx}
\usepackage{enumerate}
\usepackage{color}
\usepackage{mwe}
\usepackage{subcaption}
\usepackage{appendix}
\usepackage{float}
\usepackage{comment}
\usepackage{arydshln}
\usepackage{authblk}

\usepackage{soul}
\usepackage{xcolor}

\usepackage{tikz}
\usetikzlibrary{arrows.meta,shapes}

\usepackage{setspace}

\DeclareMathOperator{\Ber}{Ber}
\DeclareMathOperator{\logit}{logit}

\newcommand{\E}{\mathrm{E}}

\newcommand{\keywords}[1]{\textbf{\textit{Keywords---}} #1}
\newcommand{\abbreviations}[1]{\textbf{\textit{Abbreviations---}} #1}

\begin{document}

\title{Revisiting the g-null paradox}

\author[1]{Sean McGrath}
\author[2,3]{Jessica G. Young}
\author[1,3,4]{Miguel A. Hern\'an}

\affil[1]{\small Department of Biostatistics, Harvard T. H. Chan School of Public Health, Boston, MA, 02115, USA}
\affil[2]{\small Department of Population Medicine, Harvard Medical School and Harvard Pilgrim Health Care Institute, Boston, MA, 02215, USA}
\affil[3]{\small Department of Epidemiology, Harvard T. H. Chan School of Public Health, Boston, MA, 02115, USA}
\affil[4]{\small Harvard-MIT Division of Health Sciences and Technology, Cambridge, MA, 02139, USA}

\date{}
\maketitle

\begin{abstract}
  The parametric g-formula is an approach to estimating causal effects of sustained treatment strategies from observational data. An often cited limitation of the parametric g-formula is the g-null paradox: a phenomenon in which model misspecification in the parametric g-formula is guaranteed under the conditions that motivate its use (i.e., when identifiability conditions hold and measured time-varying confounders are affected by past treatment). Many users of the parametric g-formula know they must acknowledge the g-null paradox as a limitation when reporting results but still require clarity on its meaning and implications. Here we revisit the g-null paradox to clarify its role in causal inference studies. In doing so, we present analytic examples and a simulation-based illustration of the bias of parametric g-formula estimates under the conditions associated with this paradox. Our results highlight the importance of avoiding overly parsimonious models for the components of the g-formula when using this method.  
\end{abstract}

\keywords{g-null paradox, parametric g-formula, model misspecification, causal inference} \\

\abbreviations{CI confidence interval; DAG directed acyclic graph; ICE iterative conditional expectation; IP inverse probability; NICE noniterative conditional expectation NICE; ML machine learning; SE standard error; SWIG single world intervention graph}

\section{Introduction}
The g-formula identifies causal effects of sustained treatment strategies from observational data in the presence of treatment-confounder feedback \cite{Robinsfail,fitzmaurice2008estimation} under no unmeasured confounding and other assumptions \cite{Robinsfail}. A common representation of the g-formula is a non-iterative expectation weighted by the joint densities of the covariates. To obtain an estimate in a finite sample, one can first obtain estimates of each of the densities and then plug these estimates into the g-formula expression \cite{Robinsfail, WHOchap, taubman2009intervening, gfoRmula}. We refer to this estimator as a plug-in, noniterative conditional expectation (NICE), parametric g-formula estimator \cite{wen2020parametric}. For simplicity, we refer to it as the parametric g-formula in this paper.

Robins and Wasserman \cite{gnull} showed that the parametric g-formula may be guaranteed some degree of model misspecification there is treatment-confounder feedback and the sharp causal null hypothesis (i.e., the treatment has no effect on any individual's outcome at any time) is true, even if the identifying conditions hold. As a consequence, under these conditions \cite{fitzmaurice2008estimation}, a hypothesis test based on parametric g-formula estimates will falsely reject the null hypothesis of no treatment effect in large enough studies with probability approaching one \cite{gnull}. This phenomenon has been popularly referred to as the \textit{g-null paradox}.  

The existence of the g-null paradox is a potential threat to the validity of data analyses that rely on parametric g-formula estimates. There is, however, misunderstanding in the applied literature about the meaning and possible implications of the g-null paradox. Here we present analytic examples and a simulation-based illustration of the bias of the parametric g-formula under the conditions associated with this paradox.


The structure of the paper is as follows.  We first review the observed data structure, causal estimands, and the g-formula. Then, we review the example of the g-null paradox introduced in Robins and Wasserman \cite{gnull}. We clarify how model misspecification can also be guaranteed in settings other than the sharp causal null through an example. Last, we illustrate the impact of model misspecification under the conditions of the g-null paradox on bias, variance, and confidence interval coverage in a simulation study.

\section{Background}\label{sec: background}

\subsection{The observed data}
Consider an observational study with $n$ individuals for which measurements are available at regularly spaced intervals (e.g., months) denoted by $k=0,\ldots,K$ with $k=0$ the baseline interval and $K+1$ the interval in which an outcome $Y$ is of interest. For each time $k$ suppose the following are measured: $A_k$ the value of a treatment of interest (e.g., dose of a given medication) and $L_k$ a vector of covariates with $L_0$ possibly additionally containing time-fixed and pre-baseline covariates. We adopt the convention that $L_k$ precedes $A_k$ in each $k$ and use overbars to denote the history of a random variable; e.g. $\overline{A}_k := (A_0, A_1,...,A_k)$. 

The causal directed acyclic graph (DAG) in Figure \ref{fig: graphs motivating}a represents a possible data generating assumption for the observational study for the simple case of two times ($K=1$) and with the population stratified on a single level of $L_0$ (such that it can left off of the graph).  Here $U$ is an assumed unmeasured common cause of the disease outcome and a measured covariate $L_1$. For simplicity and without loss of generality, we assume that all covariates $L_k$ are discrete, and that there is no missing data, no measurement error, and no death during the study period.
\label{dags}

\subsection{The causal question}
Researchers are interested in using the data from this study to estimate the causal effect of an intervention that ensures everyone takes $150$ mg of treatment every month during the follow-up, versus 50 mg, on the mean of the outcome. The Single World Intervention Graph (SWIG) in Figure \ref{fig: graphs motivating}b is a transformation of the causal DAG under an intervention that sets treatment dose in the first two intervals to particular values $a_0$ and $a_1$, respectively \cite{richardson2013single}.  

For $a_k$ a possible level of treatment dose at $k$ and $Y^{\overline{a}_K}$ an individual's outcome if, possibly contrary to fact, the individual had adhered to a strategy assigning treatment doses $\overline{a}_K=(a_0,a_1,\ldots,a_K)$ over the follow-up, then the average causal effect is
\begin{equation}
   \E(Y^{\overline{a}_K = \overline{150}} - Y^{\overline{a}_K = \overline{50}}). \label{effect}
\end{equation}

\subsection{The g-formula}
Robins \cite{Robinsfail} showed that $\E(Y^{\overline{a}_K})$ can be identified by the \textsl{g-formula}
\begin{equation} 
    h(\overline{a}_K) = \sum_{\overline{l}_K} \E\left( Y | \overline{L}_K = \overline{l}_K, \overline{A}_K = \overline{a}_K \right) \prod_{j = 0}^K f(l_j | \overline{l}_{j - 1}, \overline{a}_{j-1}),\label{gform}
\end{equation}
where $\E\left( Y | \overline{L}_K = \overline{l}_K, \overline{A}_K = \overline{a}_K \right)$ is the mean of $Y$ among those with particular history $(\overline{l}_K, \overline{a}_K)$ and $f(l_j | \overline{l}_{j - 1}, \overline{a}_{j-1})\equiv \Pr[L_j=l_j|\overline{L}_{j - 1}=\overline{l}_{j - 1}, \overline{A}_{j-1}=\overline{a}_{j-1}]$ is the proportion of individuals with $L_j = l_j$ among those with history $(\overline{l}_{j-1}, \overline{a}_{j-1})$ and the sum is over all possible levels $\overline{l}_K$ of $\overline{L}_K$ in this population. 

\begin{figure} [ht]
\centering
{
    \centering
\begin{tikzpicture}
\begin{scope}[every node/.style={thick,draw=none}]
    \node (Label) at (-0.5, 1.25) {A)};
    \node (U) at (2,-2) {$U$};
    \node (A0) at (0,0) {$A_{0}$};
    \node (L1) at (2,0) {$L_{1}$};
    \node (A1) at (4,0) {$A_{1}$};
    \node (Y) at (6,0) {$Y$};
\end{scope}

\begin{scope}[>={Stealth[black]},
              every node/.style={fill=white,circle},
              every edge/.style={draw=black,very thick}]
    \path [->] (U) edge (L1);
    \path [->] (U) edge[bend right] (Y);
    \path [->] (A0) edge (L1);
    \path [->] (A0) edge[bend left] (A1);
    \path [->] (A0) edge[bend left] (Y);
    \path [->] (L1) edge (A1);
    \path [->] (A1) edge (Y);
\end{scope}
\end{tikzpicture} }
    \quad {
    \centering
\begin{tikzpicture}
\begin{scope}[every node/.style={thick,draw=none}]
    \node (Label) at (-0.5, 1.25) {B)};
    \node (U) at (2,-2) {$U$};
    \node (A0) at (0,0) {$A_{0}|a_0$};
    \node (L1) at (2,0) {$L_{1}^{a_0}$};
    \node (A1) at (4,0) {$A_{1}^{a_0}|a_1$};
    \node (Y) at (6,0) {$Y^{a_0, a_1}$};
\end{scope}

\begin{scope}[>={Stealth[black]},
              every node/.style={fill=white,circle},
              every edge/.style={draw=black,very thick}]
    \path [->] (U) edge (L1);
    \path [->] (U) edge[bend right] (Y);
    \path [->] (A0) edge (L1);
    \path [->] (A0) edge[bend left] (A1);
    \path [->] (A0) edge[bend left] (Y);
    \path [->] (L1) edge (A1);
    \path [->] (A1) edge (Y);
\end{scope}
\end{tikzpicture} }
    \caption{Causal graphs for the motivating example with two time intervals. Panel A illustrates a causal DAG representing an observed data generating assumption. Panel B illustrates SWIG transformation of the causal DAG under a treatment strategy $\overline{a}_1$. \label{fig: graphs motivating}}
\end{figure}
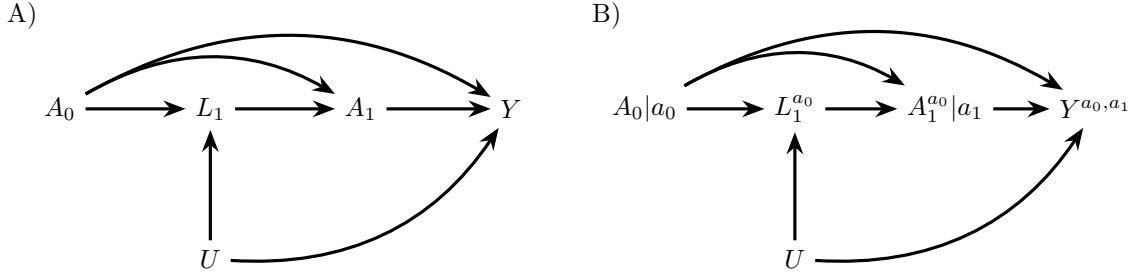

The identification of $\E(Y^{\overline{a}_1})$ by the g-formula (\ref{gform}) requires the assumptions of \textsl{sequential exchangeability, positivity, and consistency} which have been discussed at length elsewhere \cite{Robinsfail, richardson2013single, causalbook}. Sequential exchangeability, sometimes referred to as no unmeasured confounding, is encoded on the SWIG in Figure \ref{fig: graphs motivating}b by the absence of an arrow from the unmeasured $U$ into the natural values of treatment at any time \cite{richardson2013single}.


\section{The g-null paradox} \label{sec: nullparadox}

The causal effect can be estimated using the $n$ individuals in our study by estimating $h(\overline{a}_K=\overline{150})-h(\overline{a}_K=\overline{50})$.  This function may be difficult to estimate in practice when $L_k$ contains many covariates at each time $k$. The parametric g-formula is a computationally straightforward approach to estimating this function under the assumption that the components of the g-formula in (\ref{gform}) can be correctly characterized by (unsaturated) parametric models.   

However, Robins and Wasserman \cite{gnull} showed that when the following three conditions hold
\begin{itemize}
    \item Condition 1: The counterfactual mean $\E(Y^{\overline{a}_K})$ is identified by the g-formula $h(\overline{a}_K)$ in (\ref{gform}).
    \item Condition 2: The time-varying confounders $L_k$ are affected by past treatment
    \item Condition 3: The treatment has no effect on any individual's outcome at any time (i.e. the sharp null is true)
    \end{itemize}
then parametric models cannot, in general, correctly characterize the g-formula (\ref{gform}). This contradiction, which has come to be known as the \textsl{g-null paradox}, implies that, if conditions 1, 2 and 3 are true, an estimate of the effect of interest (\ref{effect}) using the parametric g-formula will be subject to some bias.

\begin{figure} [ht]
\centering
{
    \centering
\begin{tikzpicture}
\begin{scope}[every node/.style={thick,draw=none}]
    \node (Label) at (-0.5, 1.25) {A)};
    \node (U) at (2,-2) {$U$};
    \node (A0) at (0,0) {$A_{0}$};
    \node (L1) at (2,0) {$L_{1}$};
    \node (A1) at (4,0) {$A_{1}$};
    \node (Y) at (6,0) {$Y$};
\end{scope}

\begin{scope}[>={Stealth[black]},
              every node/.style={fill=white,circle},
              every edge/.style={draw=black,very thick}]
    \path [->] (U) edge (L1);
    \path [->] (U) edge[bend right] (Y);
    \path [->] (A0) edge (L1);
    \path [->] (A0) edge[bend left] (A1);
    \path [->] (L1) edge (A1);
\end{scope}
\end{tikzpicture} }
    \quad
    {
    \centering
\begin{tikzpicture}
\begin{scope}[every node/.style={thick,draw=none}]
    \node (Label) at (-0.5, 1.25) {B)};
    \node (U) at (2,-2) {$U$};
    \node (A0) at (0,0) {$A_{0}|a_0$};
    \node (L1) at (2,0) {$L_{1}^{a_0}$};
    \node (A1) at (4,0) {$A_{1}^{a_0}|a_1$};
    \node (Y) at (6,0) {$Y$};
\end{scope}

\begin{scope}[>={Stealth[black]},
              every node/.style={fill=white,circle},
              every edge/.style={draw=black,very thick}]
    \path [->] (U) edge (L1);
    \path [->] (U) edge[bend right] (Y);
    \path [->] (A0) edge (L1);
    \path [->] (A0) edge[bend left] (A1);
    \path [->] (L1) edge (A1);
\end{scope}
\end{tikzpicture} }
    \caption{Causal graphs of the original example by Robins and Wasserman \cite{gnull}. Panel A illustrates a causal DAG representing an observed data generating assumption. Panel B illustrates a SWIG transformation of the causal DAG under a treatment strategy $\overline{a}_1$. \label{fig: graphs original}}
\end{figure}
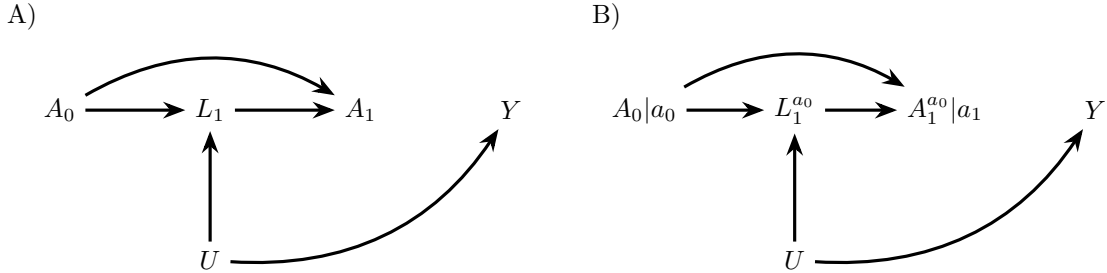

Robins and Wasserman \cite{gnull} illustrated this contradiction with an example, represented in Figure \ref{fig: graphs original}, in which Conditions 1, 2 and 3 hold. The example is a simplified version of our observational study with a null treatment effect, only two follow-up intervals ($K=1$), constant $L_0$ (thus it can be ignored) and with $L_1$ containing only one binary covariate. In this simple case, the g-formula (\ref{gform}) reduces to
\begin{equation*}
    h(a_0,a_1) = \sum_{l_1=0}^1 \E\left( Y | l_1, a_0, a_1 \right) f(l_1 | a_0).
\end{equation*}
Figure \ref{fig: graphs original} generally implies the following:
\begin{itemize}
    \item Implication 1: $\E(Y^{a_0,a_1})=h(a_0,a_1)$ (by condition 1) does not depend on $a_0$ or $a_1$ (by condition 3 -- consistent with the absence of any arrows into $Y$ except from $U$ in both panels of Figure \ref{fig: graphs original}).
    \item Implication 2: $\E\left( Y | l_1, a_0, a_1 \right)$ does depend on $l_1$ (by condition 2) -- consistent with the paths $L_1\leftarrow U\rightarrow Y$ and $L^{a_0}_1\leftarrow U\rightarrow Y$ in Figures \ref{fig: graphs original}a and \ref{fig: graphs original}b, respectively.
    \item Implication 3: $f(l_1 | a_0)$ does depend on $a_0$ (by condition 2) -- consistent with the paths $A_0\rightarrow L_1$ and $a_0 \rightarrow L_1^{a_0}$ in Figures \ref{fig: graphs original}a and \ref{fig: graphs original}b, respectively
\end{itemize}

Now suppose that a parametric model correctly characterizes the components of the g-formula:
\begin{equation*}
    \E\left( Y | l_1, a_0, a_1 \right) = g(l_1, a_0, a_1; \theta)
\end{equation*}
with $g$ a function of $(l_1, a_0, a_1)$ and a parameter vector $\theta$ and 
\begin{equation*}
    f(l_1 | a_0) = r(l_1, a_0; \beta)
\end{equation*}
with $r$ a function of $(l_1, a_0)$ and a parameter vector $\beta$ and constrained between zero and one. Given these parametric assumptions, we may replace $h(a_0, a_1)$ with the parametric g-formula
\begin{equation}
    h(a_0, a_1; \theta, \beta) = \sum_{l_1 = 0}^1 g(l_1, a_0, a_1; \theta) r(l, a_0; \beta)\label{pgform}.
\end{equation}

Robins and Wasserman \cite{gnull}, considered the following standard choices of $g$ and $r$:
\begin{align}
    g(l_1, a_0, a_1; \theta) & = \theta_0 + \theta_1 l_1 + \theta_2 a_1 + \theta_3 a_0  \label{g_exp}\\
    r(l_1 = 1, a_0; \beta) & = \frac{\exp(\beta_0 + \beta_1 a_0)}{1 + \exp(\beta_0 + \beta_1 a_0)}. \label{r_exp}
\end{align}
Plugging in these specific choices of $g$ and $r$ into (\ref{pgform}) we have
\begin{equation} \label{gform_exp}
    h(a_0, a_1; \theta, \beta) = \theta_0 + \theta_2 a_1 + \theta_3 a_0 + \frac{\theta_1 \exp(\beta_0 + \beta_1 a_0)}{1 + \exp(\beta_0 + \beta_1 a_0)}.
\end{equation}
For these choices of $g$ and $r$, it is straightforward to see that $h(a_0, a_1; \theta, \beta)$ will not depend on $(a_0, a_1)$ if and only if $\theta_2 = \theta_3 = 0$ \textsl{and} either $\theta_1 = 0$ or $\beta_1 = 0$. However, $\theta_1 = 0$ contradicts the dependence of $\E(Y | l_1, a_0, a_1)$ on $l_1$ and $\beta_1 = 0$ contradicts the dependence of $f(l_1 | a_0)$ on $a_0$.

That is, if Conditions 1, 2 and 3 hold, parametric models cannot correctly characterize the g-formula (\ref{gform}). As noted by Robins and Wasserman \cite{gnull}, adding more flexibility to models $g$ and $r$ will not remove the problem unless these parametric models are saturated. However, even in this trivialized example (with a single binary $L_1$), saturated models will be impractical if treatment at either time is continuous. Parametric models can correctly characterize the g-formula if $A$ is binary (i.e., it can take only two values) at all times and certain coefficients in models $g$ and $r$ happen to be perfect functions of others (see Appendix \ref{sec: counterexample}).

\section{Beyond the sharp causal null hypothesis} \label{sec: nonnull}
The possibility of the g-null paradox is often handled informally in practice. Investigators dismiss the g-null paradox when they find non-null effect estimates if substantive knowledge or prior studies suggested that the sharp causal null (condition 3) does not hold (e.g., see \cite{zhang2018comparing, neophytou2016occupational, garcia2018lung}) or when they find null effect estimates precisely because, despite the potential for the existence of the g-null paradox, they do find a null result (e.g., see \cite{danaei2016estimated}). 

While helpful when concerned about the existence of a non-null effect, this informal reasoning  privileges the sharp causal null and thus obscures the more general point: regardless of whether the sharp causal null holds, there may be a contradiction between the assumption that parametric models can correctly characterize the g-formula and the assumptions encoded in the causal DAG. In other words, some model misspecification may be inevitable in realistic settings. 

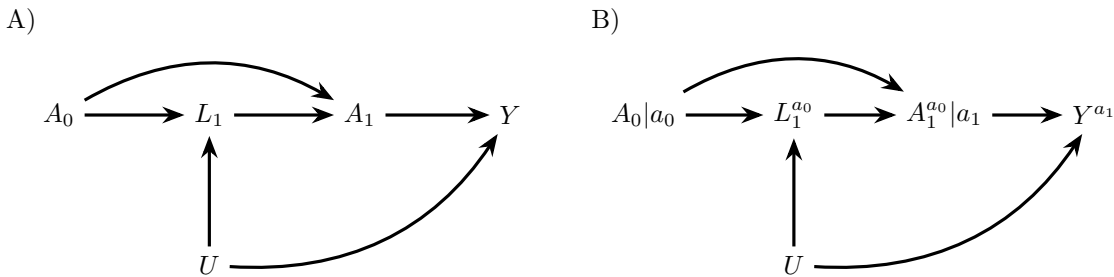
\begin{figure} [ht]
\centering
{
    \centering
\begin{tikzpicture}
\begin{scope}[every node/.style={thick,draw=none}]
    \node (Label) at (-0.5, 1.25) {A)};
    \node (U) at (2,-2) {$U$};
    \node (A0) at (0,0) {$A_{0}$};
    \node (L1) at (2,0) {$L_{1}$};
    \node (A1) at (4,0) {$A_{1}$};
    \node (Y) at (6,0) {$Y$};
\end{scope}

\begin{scope}[>={Stealth[black]},
              every node/.style={fill=white,circle},
              every edge/.style={draw=black,very thick}]
    \path [->] (U) edge (L1);
    \path [->] (U) edge[bend right] (Y);
    \path [->] (A0) edge (L1);
    \path [->] (A0) edge[bend left] (A1);
    \path [->] (L1) edge (A1);
    \path [->] (A1) edge (Y);

\end{scope}
\end{tikzpicture} }
    \quad
{
    \centering
\begin{tikzpicture}
\begin{scope}[every node/.style={thick,draw=none}]
    \node (Label) at (-0.5, 1.25) {B)};
    \node (U) at (2,-2) {$U$};
    \node (A0) at (0,0) {$A_{0}|a_0$};
    \node (L1) at (2,0) {$L_{1}^{a_0}$};
    \node (A1) at (4,0) {$A_{1}^{a_0}|a_1$};
    \node (Y) at (6,0) {$Y^{a_1}$};
\end{scope}

\begin{scope}[>={Stealth[black]},
              every node/.style={fill=white,circle},
              every edge/.style={draw=black,very thick}]
    \path [->] (U) edge (L1);
    \path [->] (U) edge[bend right] (Y);
    \path [->] (A0) edge (L1);
    \path [->] (A0) edge[bend left] (A1);
    \path [->] (L1) edge (A1);
    \path [->] (A1) edge (Y);
\end{scope}
\end{tikzpicture} }
    \caption{Causal graphs depicting the assumptions for the modification to the example by Robins and Wasserman \cite{gnull}. Panel A illustrates a causal DAG representing an observed data generating assumption. Panel B illustrates a SWIG transformation of the causal DAG under an intervention $\overline{a}_1$ \label{fig: graphs nonnull}}
\end{figure}

To see this, consider the following modification to the example in the previous section. Figure \ref{fig: graphs nonnull} represents a less restrictive assumption allowing that treatment at time $1$ may affect the outcome.  Figure \ref{fig: graphs nonnull} is in line with condition 1, 2, and a modification of condition 3 assuming that there is effect of treatment at time $0$ only (with no constraint on the effect of treatment at time $1$).  These modified conditions are encoded, for example, by the \textsl{marginal structural model} (\cite{msmref, robins2000marginal}) 
\begin{equation*}
    \E\left( Y^{a_0,a_1 }\right) = h(a_0,a_1)=\psi_0+\psi_1a_1,
\end{equation*}
Relying on the same models for $g$ and $r$ (\ref{g_exp}) and (\ref{r_exp}), we again arrive at equation (\ref{gform_exp}) and obtain the same contradiction: $h(a_0, a_1; \theta, \beta)$ will not depend on $a_0$ if and only if $\theta_3 = 0$ and either $\theta_1 = 0$ or $\beta_1 = 0$. As previously argued, the condition of $\theta_1 = 0$ or $\beta_1 = 0$ contradicts our initial assumptions that $\E(Y | l, a_0, a_1)$ depends on $l$ and $f(l | a_0)$ depends on $a_0$.

In summary, the g-null paradox is a particular instance of model misspecification that may arise when using the parametric g-formula, irrespective of whether the sharp causal null holds.

\section{Simulations} \label{sec: simulations}

We conducted numerical simulations to evaluate the impact of model misspecification on parametric g-formula estimates under the conditions of the g-null paradox.

\subsection{Simulation Design}

We considered $6$ scenarios by varying the number of follow-up time points (1, 5, or 10) and the type of treatment (continuous or binary). 

For each scenario, we simulated $250$ longitudinal data sets with $10,000$ individuals and $K+1$ time points. We evaluated the bias, standard error (SE), and confidence interval (CI) coverage of the estimator for the outcome mean under each intervention and the difference of means across interventions. The true outcome mean under each intervention was 500, and the true difference of means was 0. 
 
We first drew an unmeasured confounder ($U$) from a $\textrm{Uniform}(0,1)$ distribution. We then simulated the time-varying covariate ($L_k$) at each time $k$ ($k = 0, 1, \dots, K$) and simulated the outcome at time $K$ by
\begin{align}
    L_{k} & \sim \Ber(p= \logit^{-1}( \alpha_0 + \alpha_1 A_{k-1} + \alpha_2  U + \alpha_3 A_{k - 1} U)) \label{L_gen} \\
    Y & \sim \textrm{N}_{[0,1000]}(350 + 300 U, 50^2)
\end{align}
where the value of $\alpha_i$, $i = 0, \dots, 3$, depended on whether the treatment was continuous or discrete and $\textrm{N}_{[a, b]}(\mu, \sigma^2)$ denotes the $\textrm{N}(\mu, \sigma^2)$ distribution truncated in the interval $[a, b]$. 

In the continuous treatment scenarios, we set $(\alpha_0 = 1, \alpha_1 = -0.015, \alpha_2 = 1, \alpha_3 = 0.015)$ in (\ref{L_gen}) for the simulation of $L_k$ and simulated the time-varying treatment ($A_k$) at each time interval $k$ by
\begin{equation*}
    A_{k} \sim \textrm{N}_{[0, 200]}(80 + 0.1  A_{k-1} + 30  L_k -0.05 A_{k -1}L_k, 25^2).
\end{equation*}
In the binary treatment scenarios, we set $(\alpha_0 = 0, \alpha_1 = -2.5, \alpha_2 = 1, \alpha_3 = 2.5)$ in (\ref{L_gen}) and simulated $A_k$ by
\begin{equation*}
    A_{k} \sim \Ber(p= \logit^{-1}( -1.25 +  A_{k-1} +  L_k + A_{k-1} L_k)).
\end{equation*}
We define $L_k$ for $k = -1, \dots, -9$ as components of $L_0$ and generated them according to (\ref{L_gen}) with $A_k = A_{k-1} = 0$.

\subsubsection{Analysis of the simulated data}

We considered the interventions $\overline{a}_K = \overline{50}$ and $\overline{a}_K = \overline{150}$ in the continuous treatment scenarios and considered the interventions $\overline{a}_K = \overline{0}$ and $\overline{a}_K = \overline{1}$ in the binary treatment scenarios. 

We applied the parametric g-formula to estimate the mean of the outcome of interest at time $t = K$ and the difference of means under the above interventions. We computed 95\% CIs around all estimates using 250 bootstrap replicates.  

Observe that, by our data generating models, $Y$ and $L_k$ depend on the entire history of $L_k$ through their dependence on $U$. Because $U$ is not available to the analyst, the functional forms of the models needed for the parametric g-formula, dependent on the history of $L_k$ marginal over $U$, are therefore unknown.  We analyzed the simulated datasets in four different ways, where we modelled the history of $L_k$ with increasing flexibility.

\begin{itemize}
\item \textbf{Least Flexible}: We fit models for $L_k$ and $Y$ that include a single term for the (lagged) cumulative average value of $L_k$. In particular, we fit the following logistic model for $L_k$ and linear model for $Y$
\begin{align*}
        & \logit(\Pr[L_{k} = 1 | \overline{L}_{k-1} = \overline{l}_{k-1}, \overline{A}_{k-1} = \overline{a}_{k-1}]) = \gamma_0 + \gamma_1 a_{k-1} + \gamma_2 \frac{1}{k+9}\sum_{i = -9}^{k - 1}l_{i}\\
        & \E (Y | \overline{L}_{K} = \overline{l}_{K}, \overline{A}_{K} = \overline{a}_{K}) = \omega_0 + \omega_1 a_{K} + \omega_2 a_{K-1} + \omega_3 \sum_{i = 0}^{K - 2}a_{i} + \omega_4 \sum_{i = -9}^{K}l_{i}.
    \end{align*}
This analysis uses all the data required to satisfy the sequential exchangeability assumption, but it is expected to result in biased estimates because the parametric models will be somewhat misspecified.

\item \textbf{Moderately Flexible}: We fit models for $L_k$ and $Y$ that include terms for the two most recent lagged values of $L_k$ and a term for the lagged cumulative average value of $L_k$:
\begin{align*}
        & \logit(\Pr[L_{k} = 1 | \overline{L}_{k-1} = \overline{l}_{k-1}, \overline{A}_{k-1} = \overline{a}_{k-1}]) = \gamma_0 + \gamma_1 a_{k-1} + \gamma_2 l_{k-1} + \gamma_3 l_{k-2} + \gamma_4 \frac{1}{k+7} \sum_{i = -9}^{k-3} l_i\\
        & \E(Y | \overline{L}_{K} = \overline{l}_{K}, \overline{A}_{K} = \overline{a}_{K}) =  \omega_0 + \omega_1 a_{K} + \omega_2 a_{K-1} + \omega_3  \sum_{i = 0}^{K - 2}a_{i} + \omega_4 l_K + \omega_5 l_{K-1} + \omega_6 l_{K-2} + \omega_7 \sum_{i = -9}^{K-3}l_i.
    \end{align*}

\item \textbf{Most Flexible}: We fit models for $L_k$ and $Y$ that include a term for each lagged value of $L_k$: 
\begin{align*}
        & \logit(\Pr[L_{k} = 1 | \overline{L}_{k-1} = \overline{l}_{k-1}, \overline{A}_{k-1} = \overline{a}_{k-1}]) = \gamma_0 + \gamma_1 a_{k-1} + \sum_{i = 1}^{10} \gamma_{1+i} l_{k-i}\\
        & \E(Y | \overline{L}_{K} = \overline{l}_{K}, \overline{A}_{K} = \overline{a}_{K}) = \omega_0 + \omega_1 a_{K} + \omega_2 a_{K-1} + \omega_3  \sum_{i = 0}^{K - 2}a_{i} + \sum_{i = 0}^{10} \omega_{4 + i} l_{K-i}.
    \end{align*}

\item \textbf{Benchmark}: As a benchmark for the above three analyses, we consider an (impossible) analysis in which one has access to the unmeasured $U$ and knowledge of the functional form of the generation for $L_k$ and $Y$:
    \begin{align*}
        & \logit(\Pr[L_{k} = 1 | \overline{L}_{k-1} = \overline{l}_{k-1}, \overline{A}_{k-1} = \overline{a}_{k-1}]) = \gamma_0 + \gamma_1 a_{k-1} + \gamma_2 u + \gamma_3 a_{k-1} u \\
        & \E(Y | \overline{L}_{K} = \overline{l}_{K}, \overline{A}_{K} = \overline{a}_{K}) = \omega_0 + \omega_1 u + \omega_2 a_{K} + \omega_3 a_{K-1} + \omega_4  \sum_{i = 0}^{K - 2}a_{i}.
    \end{align*}
This analysis will be unbiased.
\end{itemize}

We applied the parametric g-formula using the \texttt{gfoRmula} \texttt{R} package \cite{gfoRmula}. The code used for all analyses is available on GitHub at https://github.com/CausalInference/NullParadox.

\subsection{Results}

\begin{figure}[H]
  \centering
  \includegraphics[width= \textwidth]{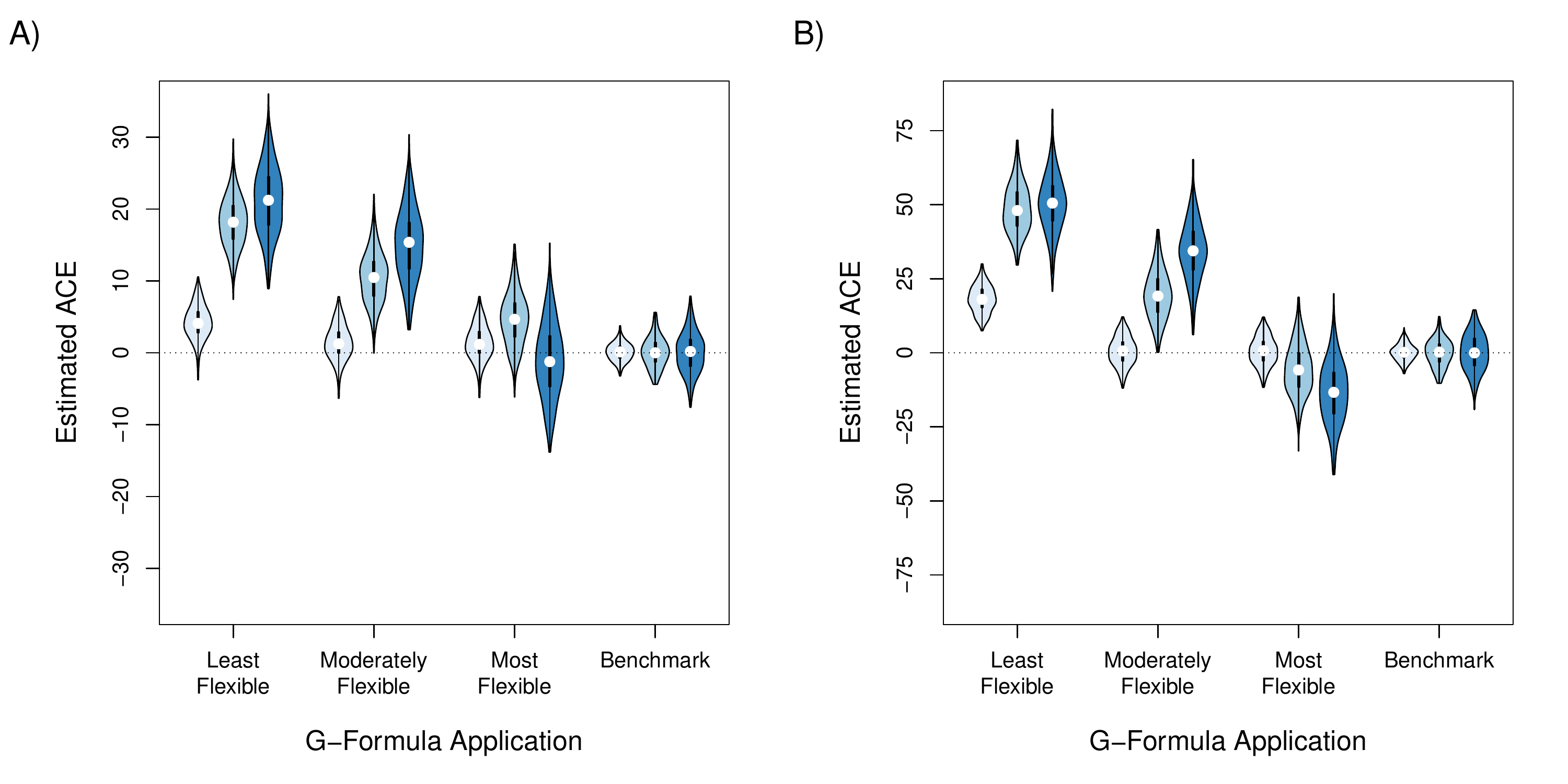}
  \caption{Violin plots illustrating the estimated average causal effect of the four applications of the parametric g-formula in the binary treatment scenarios (panel A) and continuous treatment scenarios (panel B). Darker blue shading indicates simulation settings with larger number of time points (light blue: $K=1$; blue: $K=5$; dark blue: $K=10$). The dashed line indicates the true value of the average causal effect (i.e., 0).  \label{fig: sim_res}}
\end{figure}

Figure \ref{fig: sim_res} illustrates the simulation results for the mean difference. The bias, SE, and CI coverage of the parametric g-formula are summarized in Table \ref{sim_res} in Appendix \ref{sec: additionalresults}. 

The performance of the parametric g-formula generally improved as the flexibility of the models for $L_k$ and $Y$ increased. That is, the impact of model misspecification was greatly mitigated, but not completely eliminated, by using more flexible models for the components of the parametric g-formula. For instance, at $K = 10$ in the continuous treatment scenario, the least flexible application of the parametric g-formula had a bias of 50.39, SE of 9.77, and coverage of 0.00 whereas the most flexible application had a bias of -13.53, SE of 10.38, and coverage of 0.73.

The simulation results for the counterfactual means are given in Appendix \ref{sec: additionalresults}. The same trends were observed.

\section{Discussion} \label{sec: discussion}

Our presentation clarifies that the g-null paradox of the parametric g-formula is a particular case of a general form of model misspecification that may occur even if the null does not hold. Part of the confusion surrounding the g-null paradox arose because the paradox has been traditionally discussed in the context of testing the sharp causal null hypothesis of no treatment effect. For example, Campbell and Gustafson \cite{campbellvalidity} investigated the empirical type 1 error rate based on the parametric g-formula. They did not find higher empirical type 1 error rates under the conditions of Robins and Wasserman \cite{gnull} ensuring some model misspecification compared with saturated models guaranteeing no model misspecification, although their sample sizes were smaller than those considered here.

However, when the primary goal of causal inference is estimating treatment effects, researchers are often concerned about the magnitude of bias in their estimates. Thus, one may view the g-null paradox as simply a phenomenon in which the parametric g-formula estimate of a treatment effect is guaranteed to be biased due to model misspecification, regardless of whether or not the null is true. Although not the focus of their simulation study, Murray et al. \cite{Murray2017} found the bias of the parametric g-formula to be negligible in scenarios of a null treatment effect. In contrast, our more extensive simulations illustrate that a nonnegligible amount of bias can arise in some simple scenarios even when using fairly flexible models for the components of the parametric g-formula.

Evaluating model misspecification in the parametric g-formula may be informally done by conducting sensitivity analyses under different modelling assumptions and  under different orderings of the factorization of the joint density of the confounders, comparing the parametric g-formula and the non-parametric estimate of the outcome mean/risk mean under the ``natural course"  \cite{young2011comparative}. Additionally, data analysts may consider applying approaches that are based on different algebraic representations of the the g-formula, such as inverse probability (IP) weighted estimators and the iterative conditional expectation (ICE) parametric g-formula, which rely on different modeling assumptions. ``State-of-the-art'' \cite{van2003unified, bickel1993efficient, van2002semiparametric, tsiatis2007semiparametric,  chernozhukov2018double} methods derived from the so-called efficient influence function are increasingly available \cite{lmtp_package, lmtp_paper, ltmle}.  

Yet, perhaps counterintuitively, the problem of model misspecification in the parametric g-formula is not reasonably solved by using ML algorithms to estimate the joint conditional density of the covariates and outcome. Recent simulation studies clarified that, while ``state-of-the-art'' methods can benefit from use of ML algorithms \cite{chernozhukov2018double,zivich2020machine}, the ML-based singly-robust estimators do not enjoy this benefit and may perform worse than those based on parametric models \cite{naimi2020challenges}. Therefore, we did not include such approaches in our simulations. 

In summary, because model misspecification may introduce bias in parametric g-formula estimates, it is important to avoid overly parsimonious models for the components of the g-formula when applying this method.

\section*{Acknowledgements}
The simulations in this work were run on the O2 High Performance Compute Cluster at Harvard Medical School. This work was supported by NIH grant R37 AI102634, the National Science Foundation Graduate Research Fellowship Program under Grant No. DGE1745303, National Library Of Medicine of the National Institutes of Health under Award Number T32LM012411, and Fonds de recherche du Québec-Nature et technologies B1X research scholarship. Any opinions, findings, and conclusions or recommendations expressed in this material are those of the author(s) and do not necessarily reflect the views of the funding agencies.  \\ \\
Conflict of interest: none declared.

\clearpage
\newpage
\bibliographystyle{unsrt}
\bibliography{references}

\clearpage
\newpage
\appendix

\section{A counter-example: perfect cancellation} \label{sec: counterexample}
Under Conditions 1, 2, and 3 given in the main text, the impossibility of parametric models to correctly characterize the g-formula (\ref{gform}) is expected in general. However, elucidating counter-examples exist. Specifically, suppose that only two treatment doses are prescribed in practice: 150 mg and 50 mg. In this case, redefine $A_k=1$ in the observational study if an individual receives 150 mg and $A_k=0$ if 50 mg.  

With a binary time-varying treatment, we can express the assumption of no treatment effect (condition 3) as assuming $\psi_1 = \psi_2 = \psi_3 = 0$ in the saturated marginal structural model  
\begin{equation*}
    \E\left( Y^{(a_0, a_1)}\right) = \psi_0 + \psi_1 a_1 + \psi_2 a_0 + \psi_3 a_1 a_0 
\end{equation*}
which, under identification, implies
\begin{equation*} 
    h(a_0, a_1) = \psi_0 + \psi_1 a_1 + \psi_2 a_0 + \psi_3 a_1 a_0.
\end{equation*}
We then obtain the following general expressions for $\psi_1$, $\psi_2$, and $\psi_3$ in terms of the g-formula:
\begin{align}
    \psi_1 & = h(0, 1) - h(0, 0) \label{psi1}\\
    \psi_2 & = h(1, 0) - h(0, 0) \label{psi2}\\
    \psi_3 & = h(1, 1) - h(0, 1) - h(1, 0) + h(0, 0). \label{psi3}
\end{align}

Assume the same unsaturated parametric models (\ref{g_exp}) and (\ref{r_exp}) for the g-formula in the example in Section \ref{sec: nullparadox} indexed by parameters $\theta$ and $\beta$. By plugging the expression for $h(a_0, a_1; \theta, \beta)$ in equation (\ref{gform_exp}) into the expressions for $\psi_1$, $\psi_2$, and $\psi_3$ in equations (\ref{psi1}), (\ref{psi2}), and (\ref{psi3}), we obtain the following solutions for $\psi$ in terms of $\theta$ and $\beta$,
\begin{align*}
    \psi_1 & = \theta_2 \\ 
    \psi_2 & = \theta_3 + \theta_1 \left( \frac{\exp(\beta_0 + \beta_1)}{1 + \exp(\beta_0 + \beta1)} - \frac{\exp(\beta_0)}{1 + \exp(\beta_0)} \right) \\ 
    \psi_3 & = 0.
\end{align*}
Here, we see that $\psi_1 = \psi_2 = \psi_3 = 0 $ if and only if 
\begin{itemize}
    \item $\theta_2 = \theta_3 = \theta_1 = 0$ or
    \item $\theta_2 = \theta_3 = \beta_1 = 0$ or
    \item $\theta_2 = 0$ and $\theta_3 = -\theta_1 \left( \frac{\exp(\beta_0 + \beta_1)}{1 + \exp(\beta_0 + \beta1)} - \frac{\exp(\beta_0)}{1 + \exp(\beta_0)} \right) $
\end{itemize}
The third event allows $\theta_1$ and $\beta_1$ to be non-zero (i.e., $L_1$ a time-varying confounder affected by prior treatment). Thus, there is no contradiction between the given assumptions. 

However, despite no contradiction, one might reasonably argue that the assumption of parametric models being correctly specified and that certain coefficients of these models are perfect functions of others is unreasonable. Similar arguments are given in Robins  \cite{robins2003general} and Young and Tchetgen Tchetgen \cite{young2014simulation}.

\newpage
\section{Additional simulation results} \label{sec: additionalresults}
\begin{figure}[H]
  \centering
  \includegraphics[width= \textwidth]{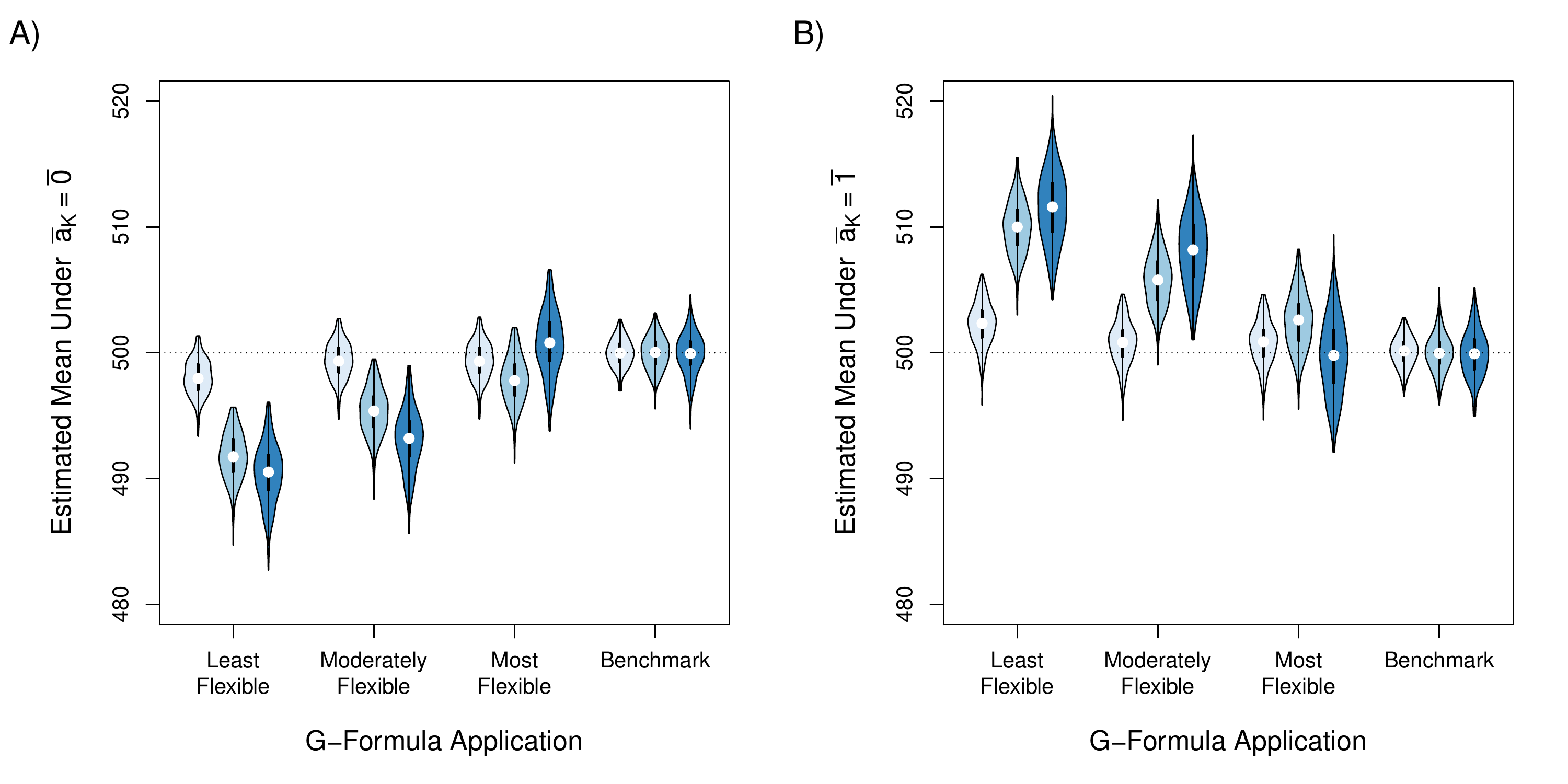}
  \caption{Violin plots illustrating the estimated counterfactual mean under the intervention $\overline{a}_K = \overline{0}$ (panel A) and intervention $\overline{a}_K = \overline{1}$ (panel B) of the four applications of the parametric g-formula in the binary treatment scenarios. Darker blue shading indicates simulation settings with larger number of time points (light blue: $K=1$; blue: $K=5$; dark blue: $K=10$). The dashed line indicates the true value of the counterfactual mean (i.e., 500).  \label{fig: sim_res_bin}}
\end{figure}

\begin{figure}[H]
  \centering
  \includegraphics[width= \textwidth]{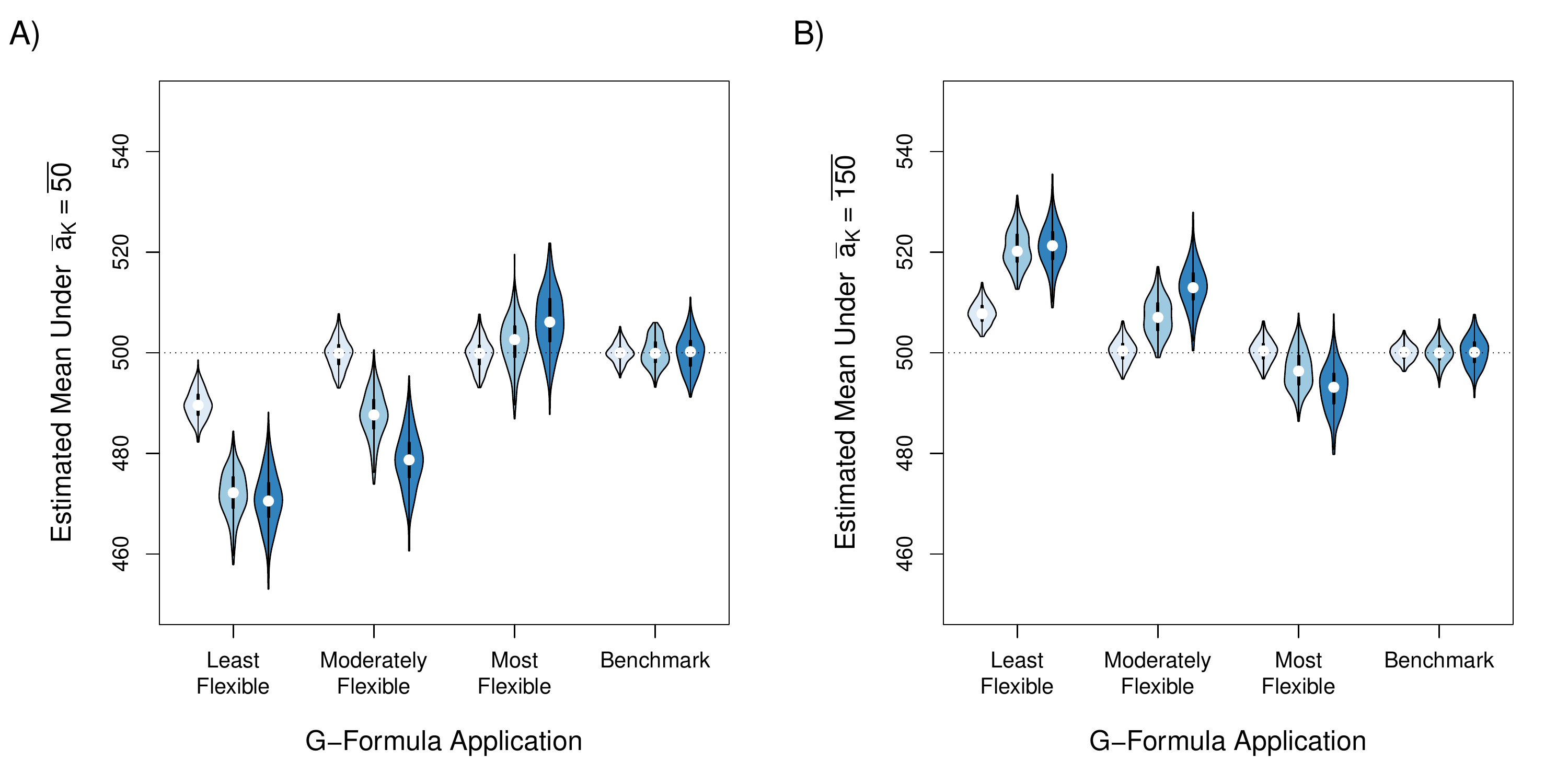}
  \caption{Violin plots illustrating the estimated counterfactual mean under the intervention $\overline{a}_K = \overline{50}$ (panel A) and intervention $\overline{a}_K = \overline{150}$ (panel B) of the four applications of the parametric g-formula in the continuous treatment scenarios. Darker blue shading indicates simulation settings with larger number of time points (light blue: $K=1$; blue: $K=5$; dark blue: $K=10$). The dashed line indicates the true value of the counterfactual mean (i.e., 500).  \label{fig: sim_res_cont}}
\end{figure}

In this section, we give additional simulation results. Table \ref{sim_res} gives the complete simulation results for the difference of means. The results for the counterfactual means in the binary and continuous treatment scenarios are summarized in Figures \ref{fig: sim_res_bin} and \ref{fig: sim_res_cont}, respectively. Tables \ref{sim_res_bin_additional} and \ref{sim_res_cont_additional} give the complete simulation results for the binary and continuous treatment scenarios, respectively.

\begin{table}[H]
\caption{Simulation results for the mean difference. The target parameter in the continuous treatment scenarios is $\E(Y^{\overline{a}_K = \overline{150}}) - \E(Y^{\overline{a}_K = \overline{50}})$ and the target parameter in the binary treatment scenarios is $\E(Y^{\overline{a}_K = \overline{1}}) - \E(Y^{\overline{a}_K = \overline{0}})$. The true value of the target parameters is 0.  \label{sim_res}}
\begin{center}
\begin{tabular}{@{\extracolsep{6pt}}lllllllll@{}}
\hline
& & \multicolumn{3}{c}{Continuous treatment scenarios} & \multicolumn{3}{c}{Binary treatment scenarios}\\
 \cline{3-5} \cline{6-9}
G-Formula Application & $K$ & Bias & SE & Coverage & Bias & SE & Coverage \\
 \hline
Least flexible & 1 & 18.19 & 4.54 & 0.01 & 4.26 & 2.39 & 0.54 \\ 
&   5 & 48.73 & 8.04 & 0.00 & 18.17 & 3.44 & 0.00 \\ 
&   10 & 50.39 & 9.77 & 0.00 & 21.14 & 4.74 & 0.00 \\ 
Moderately flexible &  1 & 0.57 & 4.76 & 0.98 & 1.42 & 2.46 & 0.86 \\ 
&   5 & 19.63 & 8.19 & 0.33 & 10.50 & 3.52 & 0.20 \\ 
&   10 & 34.43 & 10.01 & 0.07 & 15.07 & 4.88 & 0.12 \\ 
Most flexible &  1 & 0.55 & 4.76 & 0.98 & 1.42 & 2.46 & 0.86 \\ 
&   5 & -5.49 & 8.52 & 0.91 & 4.65 & 3.65 & 0.82 \\ 
&   10 & -13.53 & 10.38 & 0.73 & -1.05 & 5.13 & 0.94 \\  \hdashline
Benchmark & 1 & 0.10 & 2.77 & 0.91 & 0.11 & 1.22 & 0.96 \\ 
&   5 & -0.09 & 4.52 & 0.94 & 0.05 & 2.10 & 0.93 \\ 
&   10 & 0.10 & 6.24 & 0.94 & 0.03 & 2.81 & 0.93 \\ 
\hline
\end{tabular}
\end{center}
\end{table}

\newpage

\newpage
\begin{table}[H]
\small
\caption{Simulation results for the counterfactual means in the binary treatment scenarios. The true value for all target parameters is 500.  \label{sim_res_bin_additional}}
\begin{center}
\begin{tabular}{llllll}
\hline
Target parameter & G-Formula Application & $K$ & Bias & SE & Coverage  \\
 \hline
$\E(Y^{\overline{a}_K = \overline{0}})$ & Least flexible  & 1 & -1.96 & 1.42 & 0.66 \\ 
& &   5 & -8.19 & 1.86 & 0.00 \\ 
& &   10 & -9.59 & 2.36 & 0.01 \\ 
& Moderately flexible & 1 & -0.61 & 1.46 & 0.90 \\ 
& &   5 & -4.66 & 1.87 & 0.35 \\ 
& &   10 & -6.92 & 2.41 & 0.13 \\ 
& Most flexible & 1 & -0.62 & 1.46 & 0.90 \\ 
& &   5 & -2.11 & 1.89 & 0.84 \\ 
& &   10 & 0.76 & 2.50 & 0.93 \\ 
& Benchmark & 1 & -0.03 & 1.15 & 0.93 \\ 
& &   5 & -0.03 & 1.33 & 0.96 \\ 
& &   10 & -0.10 & 1.63 & 0.94 \\ 
$\E(Y^{\overline{a}_K = \overline{1}})$ & Least flexible  & 1 & 2.29 & 1.70 & 0.73 \\ 
& &   5 & 9.97 & 2.07 & 0.01 \\ 
& &   10 & 11.55 & 2.80 & 0.01 \\ 
& Moderately flexible & 1 & 0.80 & 1.72 & 0.91 \\ 
& &   5 & 5.84 & 2.14 & 0.30 \\ 
& &   10 & 8.15 & 2.89 & 0.17 \\ 
& Most flexible & 1 & 0.80 & 1.73 & 0.91 \\ 
& &   5 & 2.53 & 2.24 & 0.81 \\ 
& &   10 & -0.29 & 3.03 & 0.94 \\ 
& Benchmark & 1 & 0.09 & 1.21 & 0.96 \\ 
& &   5 & 0.02 & 1.53 & 0.94 \\ 
& &   10 & -0.07 & 1.86 & 0.93 \\ 
\hline
\end{tabular}
\end{center}
\end{table}

\newpage

\newpage
\begin{table}[H]
\caption{Simulation results for the counterfactual means in the continuous treatment scenarios. The true value for all target parameters is 500.  \label{sim_res_cont_additional}}
\begin{center}
\begin{tabular}{llllll}
\hline
Target parameter & G-Formula Application & $K$ & Bias & SE & Coverage  \\
 \hline
$\E(Y^{\overline{a}_K = \overline{50}})$ & Least flexible  & 1 & -10.32 & 2.86 & 0.04 \\ 
& &   5 & -28.03 & 4.60 & 0.00 \\ 
& &   10 & -29.22 & 5.59 & 0.00 \\ 
& Moderately flexible & 1 & -0.24 & 2.93 & 0.97 \\ 
& &   5 & -12.48 & 4.72 & 0.25 \\ 
& &   10 & -21.36 & 5.70 & 0.04 \\ 
& Most flexible & 1 & -0.24 & 2.94 & 0.97 \\ 
& &   5 & 2.08 & 4.89 & 0.92 \\ 
& &   10 & 6.36 & 5.93 & 0.80 \\ 
& Benchmark & 1 & -0.01 & 1.87 & 0.92 \\ 
& &   5 & 0.08 & 2.74 & 0.96 \\ 
& &   10 & 0.01 & 3.59 & 0.94 \\ 
$\E(Y^{\overline{a}_K = \overline{150}})$ & Least flexible  & 1 & 7.87 & 2.15 & 0.06 \\ 
& &   5 & 20.70 & 3.68 & 0.00 \\
& &   10 & 21.17 & 4.38 & 0.00 \\ 
& Moderately flexible & 1 & 0.32 & 2.29 & 0.98 \\
& &   5 & 7.15 & 3.72 & 0.48 \\ 
& &   10 & 13.07 & 4.52 & 0.15 \\ 
& Most flexible & 1 & 0.32 & 2.28 & 0.98 \\ 
& &   5 & -3.41 & 3.88 & 0.88 \\ 
& &   10 & -7.17 & 4.67 & 0.66 \\ 
& Benchmark & 1 & 0.10 & 1.62 & 0.92 \\ 
& &   5 & -0.01 & 2.20 & 0.94 \\ 
& &   10 & 0.11 & 2.95 & 0.93 \\ 
\hline
\end{tabular}
\end{center}
\end{table}

\newpage

\end{document}